\documentclass[reprint,prb,amsmath,amssymb,superscriptaddress]{revtex4-2}

\usepackage{esvect}
\usepackage{amsfonts, amsmath, amsthm, amssymb}
\usepackage{graphicx}
\usepackage{dcolumn}
\usepackage{bm}
\usepackage{xcolor}
\usepackage{float}
\usepackage[normalem]{ulem} 
\usepackage[export]{adjustbox}
\usepackage[ colorlinks,
    linkcolor={blue},
    citecolor={blue},
    urlcolor={blue}]{hyperref}

\newcommand\eqt{\hspace{0.17em}{=}\hspace{0.17em}}
\newcommand\coloneqqt{\hspace{0.17em}{:=}\hspace{0.17em}}
\newcommand\gt{\hspace{0.23em}{>}\hspace{0.12em}}
\newcommand\intext{\hspace{0.15em}{\in}\hspace{0.15em}}
\newcommand{\etafreq}{\eta_{\vv{\gamma},\vv{\omega}} }
\newcommand{\etafreqminimaxgrid}{\eta_{\freqminimaxgridstilde} }
\newcommand{\etatime}{\hat{\eta}_{\vv{\sigma},\vv{\tau}} }

\newcommand{\freqminimaxgrids}{\{\gks,\wks\}_{k=1}^N}
\newcommand{\timeminimaxgrids}{\{\sis,\tis\}_{i=1}^N}
\newcommand{\freqminimaxgridstilde}{\{\tgks,\twks\}_{k=1}^N}
\newcommand{\timeminimaxgridstilde}{\{\tsis,\ttis\}_{i=1}^N}
\newcommand{\freqgrids}{\{\gamma_k,\omega_k\}_{k=1}^N}
\newcommand{\timegrids}{\{\sigma_i,\tau_i\}_{i=1}^N}
\newcommand{\argminfreq}{\underset{\freqgrids}{\text{arg min}}}
\newcommand{\argmintime}{\underset{\timegrids}{\text{arg min}}}
\newcommand{\emin}{\epsilon_\text{min}}
\newcommand{\emax}{\epsilon_\text{max}}
\newcommand{\tgks}{\tilde{\gamma}_k}
\newcommand{\twks}{\tilde{\omega}_k}
\newcommand{\tsis}{\tilde{\sigma}_i}
\newcommand{\ttis}{\tilde{\tau}_i}
\newcommand{\gks}{\gamma_k^*}
\newcommand{\wks}{\omega_k^*}
\newcommand{\sis}{\sigma_i^*}
\newcommand{\tis}{\tau_i^*}
\newcommand{ \Fhe}{\hat{F}^\text{even}}
\newcommand{ \Fho}{\hat{F}^\text{odd}}
\newcommand{ \Fe}{F^\text{even}}
\newcommand{ \Fo}{F^\text{odd}}
\newcommand{\maxxoneR}{\max_{x \in [1,R]}}

\newcommand{\indk}{^{(k)}}
\newcommand{\bX}{\mathbf{X}\indk}

\usepackage{yfonts}
\begin{document}

\title{Validation of the GreenX library time-frequency component for efficient \textit{GW} and RPA calculations}

\author{Maryam Azizi}
\affiliation{Université\ Catholique\ de\ Louvain,\ Louvain-la-Neuve, Belgium}

\author{Jan Wilhelm}
\affiliation{Institute of Theoretical Physics and Regensburg Center for Ultrafast Nanoscopy (RUN), University\ of Regensburg,\ Regensburg,\ Germany}

\author{Dorothea Golze}
\affiliation{Faculty for Chemistry and Food Chemistry, Technische Universit\"at Dresden, 01062 Dresden, Germany}

\author{Francisco A. Delesma}
\affiliation{Faculty for Chemistry and Food Chemistry, Technische Universit\"at Dresden, 01062 Dresden, Germany}
\affiliation{Department of Applied Physics, Aalto University, FI-02150 Espoo, Finland}

\author{Ramón L. Panadés-Barrueta}
\affiliation{Faculty for Chemistry and Food Chemistry, Technische Universit\"at Dresden, 01062 Dresden, Germany}

\author{Patrick Rinke}
\affiliation{Department of Applied Physics, Aalto University, FI-02150 Espoo, Finland}

\author{Matteo Giantomassi}
\affiliation{Université\ Catholique\ de\ Louvain,\ Louvain-la-Neuve, Belgium}

\author{Xavier Gonze}
\affiliation{Université\ Catholique\ de\ Louvain,\ Louvain-la-Neuve, Belgium}

\date{\today}

\begin{abstract} 
Electronic structure calculations based on many-body perturbation theory (e.g. \textit{GW} or the random-phase approximation (RPA)) require function evaluations in the complex time and frequency domain, for example inhomogeneous Fourier transforms or analytic continuation from the imaginary axis to the real axis.
For inhomogeneous Fourier transforms, the time-frequency component of the GreenX library provides time-frequency grids that can be utilized in low-scaling RPA and \textit{GW} implementations.  In addition, the adoption of the compact frequency grids provided by our library also reduces the computational overhead in RPA implementations with conventional scaling.
In this work, we present low-scaling \textit{GW} and conventional RPA benchmark calculations using the GreenX grids with different codes (FHI-aims, CP2K and ABINIT) for molecules, two-dimensional materials and solids.
Very small integration errors are observed when using 30 time-frequency points for our test cases, namely $<10^{-8}$\,eV/electron for the RPA correlation energies, and $\le 10$~meV for the $GW$ quasiparticle energies.
\end{abstract}

\maketitle

\section{\label{sec:introduction}Introduction}
\textit{GW} calculations~\cite{hedin1965new} have become part of the standard toolbox in computational condensed-matter physics for the calculation of photoemission and optoelectronic spectra of molecules and solids~\cite{golze2019gw,reining2018gw,stankovski2011g,li2005quasiparticle,bruneval2008accurate,nabok2016accurate}. 
Recent highlights include a wide array of advancements, such as the application of $GW$ to deep core excitations~\cite{aoki2018accurate,golze2018core,golze2020accurate,keller2020relativistic,zhu2021all,Mejia2021scalable,Mejia2022basis,li2022benchmark,panades2023},
$GW$ studies of two-dimensional (2D) materials~\cite{Molina2013,qiu2016screening,Gjerding2021recent,Rasmussen2021,Mitterreiter2021,Lin2021,Guandalini2023efficient,graml2023lowscaling} and metal-halide perovskites~\cite{Filip2014,Cho2019,Biega2023,McArthur2023,Biffi2023,Leppert2024}. Additionally, research has delved into exploring excited-state potential energy surfaces from \textit{GW}+Bethe-Salpeter~\cite{Caylak2021,Berger2021,knysh2023excited} and applying the \textit{GW} methodology in magnetic fields~\cite{holzer2019GW,franzke2022nmr,holzer2023practical}.  Relativistic \textit{GW} schemes with two-compoment spinors have recently gained attention to treat spin-orbit coupling~\cite{Kuehn2015one,holzer2019ionized,foerster2023twocomponent,kehry2023}. Furthermore, machine learning models  have been developed for quasiparticle energies~\cite{stuke2020atomic,Westermayr2021,Golze2022,Fediai2023,Mondal2023,Zauchner2023,Venturella2024} and studies have investigated electron dynamics from Green's functions~\cite{attaccalite2011realtime,jiang2021realtime,chan2021giant,Schluenzen2020achieving,Perfetto2022realtime,Tuovinen2020comparing,tuovinen2021electronic,tuovinen2023,Pavlyukh2023}. Additionally, there have been focused studies related to $GW$ itself, like
 benchmarking the accuracy of the $GW$ method
\cite{bruneval2021gw,Orlando2023,Marek2024,Ammar2024} and  benchmarking the numerical precision of $GW$ implementations~\cite{van2015gw,maggio2017gw100,setten2017,gao2019real,govoni2018gw100,foerster2021gw100,rangel2020reproducibility}.

For the calculation of total energies, the random phase approximation (RPA)~\cite{bohm1953collective,gell1957correlation} offers several appealing features. It includes long-range dispersion interactions and dynamic electronic screening, that are absent from conventional density functionals. Despite its large computational cost, RPA has been applied to a wide range of systems,  from 0 to 3 dimensions \cite{Eshuis2012,Ren2012,harl2010assessing,furche2001molecular,fuchs2002accurate,scuseria2008ground,spadetto2023toward}. Recent highlights include RPA forces \cite{Furche/RPA-forces:2014,Kresse/RPAforces:2017,Ren/RPA-forces:2022,Ochsenfeld/RPA-forces:2022,Bussy2023,Stein2024}, RPA-based interatomic potentials \cite{Kresse/RPA-MLIP:2022,Kresse/RPA-MLIP:2023,Kresse/RPA-MLIP2:2023} and applications to complex systems such as phase transitions in solid hydrogen \cite{Casula/hydrogen:2022} and cesium-lead-triiodide \cite{CsPbI3_phase_transition:2022}, oxychlorinated platinum complexes \cite{Hellier/etal:2023}, chromophores \cite{Foerster/chromophores:2023}, and perhalogenated benzene clusters \cite{Tyagi/etal:2023}.

Several algorithmic bottlenecks, however, render \textit{GW} and RPA calculations challenging, especially for complex or disordered systems with large simulations cells. In their conventional 
implementations,  the computational cost increases with the fourth power of the system size $N_\text{at}$, $\mathcal{O}(N_\text{at}^4)$. As a consequence, conventional \textit{GW} calculations are usually limited to systems with a few hundred of atoms~\cite{wilhelm2016gw, stuke2020atomic}. 
There are many approaches to make larger system sizes tractable, which include massively parallel implementations over physically motivated approximations to novel numerical methods. Efficient parallelization schemes were developed for execution on more than 10,000 CPU cores~\cite{govoni2015large,wilhelm2016gw,Kim2019scalable,Sangalli2019many,del2019large} and first algorithms have been already proposed for the new generation of heavily GPU-based (pre)exascale supercomputers~\cite{delben2020accelerating,yu2020gpu,yeh2022fully}. An example for more physically motivated developments are $GW$ embedding schemes, where a small part of the system is calculated at the $GW$ level and the surrounding medium is treated at a lower level of theory~\cite{duchemin2016combining,Li2016,Li2018,toelle2021subsystem,amblard2023manybody}. 

In order to reduce the computational cost of \textit{GW} and RPA calculations, an alternative path relies on low-scaling algorithms that allow one to tackle larger and more realistic systems.
Low-scaling $GW$ methods can be constructed using space-local representations and imaginary time-frequency transforms~\cite{rojas1995space} so that cubic scaling of the computational cost in the system size~$N_\text{at}$ is achieved, $\mathcal{O}({N_\text{at}^3})$, instead of quartic scaling $\mathcal{O}({N_\text{at}^4})$ of conventional $GW$ algorithms~\cite{wilhelm2016gw}.

Several cubic scaling algorithms have been recently implemented,
e.g.~using a plane-wave basis set and real-space grids~\cite{kaltak2014low,kaltak2014cubic,liu2016cubic,Yeh2024} or
 using atom-orbital-like basis functions~\cite{Wilhelm2018,wilhelm2021low,graml2023lowscaling,duchemin2021cubic,foerster2020low,foerster2021loworder,foerster2022quasiparticle,foerster2023twocomponent}. 
All of these $GW$ algorithms have in common that three Fourier transforms  between imaginary time and imaginary frequency and vice versa need to be performed numerically. 
The total computational cost increases linearly with the number of time and frequency grid points which makes it attractive to construct numerically accurate grids with a minimal number of grid points. 
One approach is to construct imaginary time and frequency grids using the \textit{minimax} approach~\cite{kaltak2014low,liu2016cubic}.
We have recently published minimax time-frequency grids in the open-source GreenX library~\cite{GreenX,Azizi2023} which is interfaced to the electronic structure codes FHI-aims~\cite{Blum2009}, CP2K~\cite{Kuehne2020} and ABINIT~\cite{Gonze2020a,Gonze2020b}.

In this work, we use minimax grids from the GreenX library~\cite{GreenX,Azizi2023} together with the three codes to perform accuracy benchmark calculations for $GW$ quasiparticle energies and RPA correlation energies. Our goal is to examine the validity of the grids for a wide range of finite and extended systems. We aim to demonstrate that the GreenX grids are reliable for low-scaling algorithms, and that also conventional RPA implementations with quartic scaling benefit from our library.

The article is organized as follows:
In Sec.~\ref{acc_meth}, we discuss the low-scaling $GW$ and RPA algorithm in imaginary time and imaginary frequency. 
Sec.~\ref{sec:mini_max_theory} describes the minimax formalism for creating imaginary-time and imaginary-frequency grids. 
We finally present low-scaling $GW$ and conventional-scaling RPA benchmark calculations on molecules, 2D materials and solids, where computational details are given in Sec.~\ref{computational_details}. We present and discuss the results in Sec.~\ref{sec:test} and provide our conclusions in Sec~\ref{sec:sum}. 

\section{\label{acc_meth}RPA and \textit{GW} methodology  in imaginary time and frequency}
 In this section, we define the Green's functions, susceptibility and Fourier transformations from imaginary time to imaginary frequency and vice versa which are then used to compute the RPA correlation energy and \textit{GW} quasiparticle energies.

Within the adiabatic-connection fluctuation-dissipation formalism~\cite{gunnarsson1976exchange,dahlen2006variational,dobson1994topics,Eshuis2012,Ren2012}, the $\mathrm{RPA}$ correlation energy can be written as
\begin{equation}\label{eq:rpa}
    E_c^{\mathrm{RPA}}= \int_{0}^{\infty} \frac{d\omega}{2\pi} \mathrm{Tr} \{ \mathrm{ln}[1-\chi(i\omega)V] + \chi(i\omega)V \},
\end{equation}
where $\chi$ stands for the RPA independent-particle susceptibility (or interacting density response) evaluated on the imaginary frequency axis ($i\omega$ with $\omega$ being real) and $V$ denotes the frequency-independent Coulomb interaction.  The spatial arguments have been omitted for brevity. 
The RPA susceptibility is also an ingredient for the $GW$ self-energy.

The RPA susceptibility can be computed from the independent-particle (irreducible) non-interacting susceptibility, $\chi^0$, through Dyson's equation
\begin{equation}\label{eq:Dyson}
   \chi(i\omega)=\chi^0(i\omega)+\chi^0(i\omega)V\chi(i\omega)\,.
\end{equation}
For $GW$, the screened Coulomb interaction $W$ is then needed, followed by its frequency-convolution with the Green function $G$, as we will outline in more detail below.

In the Adler-Wiser formula~\cite{Adler1962,Wiser1963}, the non-interacting susceptibility in the imaginary-frequency domain can be obtained as follows 
\begin{eqnarray}\label{susceptibility}
\chi^0\left( \mathbf{r}, \mathbf{r'}, i\omega \right) &= &\sum_{j}^\text{occ} \sum_a^\text{unocc} \psi^*_a\left(\mathbf{r'}\right)\psi_j\left(\mathbf{r'}\right)\psi^*_j\left(\mathbf{r}\right)\psi_a\left(\mathbf{r}\right) \nonumber\\
&\times& \frac{2\left(\epsilon_j - \epsilon_a\right)}{\omega^2 + \left(\epsilon_j - \epsilon_a\right)^2},   
\end{eqnarray}
where indices $j$ and $a$ refer to occupied and unoccupied Kohn-Sham (KS) or Hartree-Fock states ($\psi(\mathbf{r})$) with energies 
$\epsilon$.  The computational cost required to compute Eq.~(\ref{susceptibility}) in the frequency domain scales as $\mathcal{O}(N_\text{at}^4)$, since the number of states ($j$ and $a$ indices) and discretized real space points 
($\mathbf{r}$ and $\mathbf{r'}$) scale linearly with system size $N_\text{at}$. This is a significant bottleneck, which can be reduced by invoking the low-scaling space-time approach~\cite{rojas1995space}. 

By Fourier transforming Eq.~(\ref{susceptibility}) in imaginary time ($i\tau$), the two summations decouple, 
\begin{eqnarray}\label{susceptibility_low}
    \hat{\chi}^0\left( \mathbf{r}, \mathbf{r'}, i\tau \right) &=& -\sum_j^\text{occ}\psi_j\left(\mathbf{r'}\right)\psi^*_j\left(\mathbf{r}\right)e^{\epsilon_j|\tau|}\nonumber\\
    &\times& \sum_a^\text{unocc}\psi^*_a\left(\mathbf{r'}\right)\psi_a\left(\mathbf{r}\right)e^{-\epsilon_a|\tau|}.  
\end{eqnarray}
A circumflex accent is used to denote Fourier transformed functions of the imaginary time.
For completeness, we mention here that Eq.~\eqref{susceptibility_low} is equivalent to expressing $\chi^0$ as a product of two non-interacting Green's functions
\begin{equation}\label{green}
    \hat{\chi}^0\left( \bold{r}, \bold{r'}, i\tau \right) = \hat{G}\left( \bold{r}, \bold{r'}, i\tau \right)\hat{G}^*\left( \bold{r}, \bold{r'}, -i\tau \right),
\end{equation}
with 
\begin{align}
    \hat{G}\left( \bold{r}, \bold{r'}, i\tau \right) &= \sum_j^{\mathrm{occ}} \psi_j(\bold{r})\psi^*_j(\bold{r'}) e^{-\epsilon_j \tau}\hspace{2.5em} (\tau < 0),
    \\
    \hat{G}\left( \bold{r}, \bold{r'}, i\tau \right) &= -\sum_a^{\mathrm{unocc}} \psi_a(\bold{r})\psi^*_a(\bold{r'}) e^{-\epsilon_a \tau}\hspace{1em}  (\tau > 0).
\end{align}
The computational cost of the Green's function scales cubically in $N_\text{at}$ and so does $\hat{\chi}^0$. The space-time approach therefore saves one order in $N_\text{at}$ compared to the aforementioned Adler-Wiser formalism.

With Fourier transformations $\chi^0$ and $\chi$ can be switched between (imaginary) time and frequency:
\begin{eqnarray}
    f\left( \bold{r}, \bold{r'}, i\omega \right) = \int_{-\infty}^{+\infty} e^{-i \omega \tau} \hat{f}\left( \bold{r}, \bold{r'}, i\tau \right) d\tau,\nonumber \\
    \hat{f}\left( \bold{r}, \bold{r'}, i\tau \right) = \frac{1}{2\pi}\int_{-\infty}^{+\infty} e^{i \omega \tau} f\left( \bold{r}, \bold{r'}, i\omega \right) d\omega.
\end{eqnarray}
Since both susceptibilities are even functions in frequency and time ($\chi^0(i\omega) = \chi^0(-i\omega)$ and $\hat{\chi}^0(i\tau) = \hat{\chi}^0(-i\tau)$),
these transforms simplify to cosine transformations,
\begin{eqnarray}
    \chi^0\left( \bold{r}, \bold{r'}, i\omega \right)& =& \int_{-\infty}^{+\infty} e^{i \omega \tau} \;\hat{\chi}^0\left( \bold{r}, \bold{r'}, i\tau \right) d\tau\nonumber\\
    & =&\int_{-\infty}^{+\infty} (\mathrm{cos}(\omega \tau) + i \mathrm{sin}(\omega \tau))\; \hat{\chi}^0\left( \bold{r}, \bold{r'}, i\tau \right) d\tau\nonumber\\
    & =&\int_{-\infty}^{+\infty}\mathrm{cos}(\omega \tau)\;\hat{\chi}^0\left( \bold{r}, \bold{r'}, i\tau \right) d\tau\nonumber\\
    & = & 2\int_{0}^{+\infty}\mathrm{cos}(\omega \tau)\;\hat{\chi}^0\left( \bold{r}, \bold{r'}, i\tau \right) d\tau.\label{susceptibility_cosine}
\end{eqnarray}

After a Fourier transforming $\chi$ to imaginary frequency, the dielectric function can be calculated in the imaginary-frequency domain as
\begin{eqnarray}\label{diel}
    \epsilon\left( \bold{r}, \bold{r'}, i\omega \right)& = & \delta\left( \bold{r}, \bold{r'}\right)\nonumber\\
    & - &\int d \bold{r''} V\left( \bold{r}, \bold{r''} \right)\chi^0\left( \bold{r''}, \bold{r'}, i\omega \right).
\end{eqnarray}
The screened Coulomb interaction $W$ is then given by
\begin{equation}
    W\left(\bold{r}, \bold{r'}, i\omega\right) = \int d\bold{r''} \epsilon^{-1}\left( \bold{r}, \bold{r''}, i\omega\right)V\left( \bold{r''},  \bold{r'}  \right)
\end{equation}
and cosine transformed into the time domain, 
\begin{align}
     W(\bold{r}, \bold{r'}, i\tau)  & = \frac{1}{2\pi} \int_{-\infty}^{+\infty} e^{i \omega \tau}\,  W(\bold{r}, \bold{r'}, i\omega)  \,  d\omega \,,\nonumber
     \\
     &=  \frac{1}{\pi} \int_{0}^{+\infty} \cos(\omega \tau)\,  W(\bold{r}, \bold{r'}, i\omega)  \,  d\omega \,.
     \label{W_cosine}
\end{align}
The self-energy follows as a product with the Green's function
\begin{align}
    \hat{\Sigma} \left( \bold{r}, \bold{r'}, i\tau \right) = i \hat{G}( \bold{r}, \bold{r'}, i\tau )\hat{W}( \bold{r}, \bold{r'}, i\tau ).
\end{align}
The quasiparticle energies are then obtained from the matrix elements of the self-energy with respect to the single-particle wave functions of the corresponding states after  Fourier transforming the self-energy from imaginary time to imaginary frequency.  Since the self-energy is neither an odd nor an even function, both cosine and sine transformations are needed~\cite{liu2016cubic},
\begin{eqnarray}\label{self-omega}
 \Sigma(i\omega) &=& -\int_{\infty}^{\infty} d \tau \,\hat G(i\tau) \hat W (i\tau) e^{i\omega t} \\[0.2em] &=& 2 \int_0^{\infty} d \tau \,\hat{\Sigma}^c(i\tau) \cos(\omega \tau) \\[0.2em] &+& \,2i \int_0^{\infty} d \tau\, \hat{\Sigma}^s(i\tau) \sin(\omega \tau),
\end{eqnarray}
where
\begin{eqnarray}
 \hat{\Sigma}^c(i\tau) = -\frac{1}{2}[\hat{G}(i\tau)+\hat{G}(-i\tau)]\hat{W}(i\tau), \\[0.5em]
 \hat{\Sigma}^s(i\tau) = -\frac{1}{2}[\hat{G}(i\tau)-\hat{G}(-i\tau)]\hat{W}(i\tau).
\end{eqnarray}
This is the last step in the low-scaling algorithm which will then be followed by the calculation of the 
quasiparticle energy using analytic continuation~\cite{golze2019gw,liu2016cubic,wilhelm2021low}.

\section{\label{sec:mini_max_theory}Minimax time and frequency integration grids}

\subsection{\label{sec:mini_max}Analytical form of functions in time and frequency}

In Sec.~\ref{acc_meth} we have discussed that the computation of the susceptibility in the imaginary time domain
scales favorably. 
However, the subsequent time-frequency Fourier transform of such a function is challenging if performed numerically.
Eq.~(\ref{susceptibility}) in the frequency domain is the exact (cosine) Fourier transform of 
Eq.~(\ref{susceptibility_low}), in the time domain. 
In a numerical approach, Eq.~(\ref{susceptibility_low}) has to be evaluated for a finite set of time points, where the set should be as small as possible for an efficient algorithm.
However, the functions introduced in the previous section in the imaginary time and frequency domains  usually have long tails and very localized features. 
As such, a usual Fast Fourier Transform, with homogeneously spaced integration grids, would need numerous sampling points. 
Instead, a nonuniform Fourier transform  (actually cosine and sine transforms) would yield a reduction by more than one order of magnitude in the computational requirements (both CPU time and memory).

The problem is thus to find the ``best set'' of time  points, and associated weights, to discretize the integral in Eq.~(\ref{susceptibility_cosine}).
Similarly, the inverse transform Eq.~\eqref{W_cosine} from frequency to time has to be performed on an equivalently ``best set'' of frequency points.
One option for defining a ``best set" of points and weights is to exploit the functional form of the frequency dependence of Eq.~(\ref{susceptibility}), and the  time dependence of Eq.~(\ref{susceptibility_low}).
For simplicity~\cite{kaltak2014low}, we single out the frequency and time dependence of Eqs.~\eqref{susceptibility} and~\eqref{susceptibility_low}
\begin{align}\label{pol_freq}
    \chi(i\omega) &= \sum_{\mu} \chi_{\mu} \phi_{\omega}(x_\mu),
\\
\label{pol_time}
    \hat{\chi}(i\tau) &= \sum_{\mu} \chi_{\mu} \hat{\phi}_{\tau}(x_\mu),
\end{align}
thanks to the auxiliary functions 
\begin{align}\label{axi_omega}
    \phi_{\omega}(x) &:= \frac{2x}{x^2+\omega^2}
\\[0.5em] \label{axi_tau}
    \hat{\phi}_{\tau}(x) &:= e^{-x|\tau|}\,,
\end{align}
where $\mu$ in Eqs.~\eqref{pol_freq} and~\eqref{pol_time} runs over the occupied-state index $j$ and the unoccupied-state index~$a$, $x_{\mu}$ is the energy difference between occupied and unoccupied states ($x_{\mu}=\epsilon_{a}-\epsilon_{j} > 0$) and $\chi_{\mu}= \psi^*_a(\bold{r'})\psi_j(\bold{r'})\psi^*_j(\bold{r})\psi_a(\bold{r})$ denotes the elements of the susceptibility matrix in the transition space. 
Note that \mbox{$\epsilon_\text{min} \leq x_{\mu} \leq \epsilon_\text{max}$}, with $\epsilon_\text{min}$ being the energy or band gap and $\epsilon_\text{max}$ the maximum energy difference.

\subsection{Time and frequency integration for the direct MP2 correlation energy}
We make use of the functional forms expressed in Eqs.~\eqref{pol_freq} and \eqref{pol_time} in the following way~\cite{kaltak2014low}: In Eq.~\eqref{eq:rpa}, for the RPA correlation energy, the function $\ln(1-x)+x=-x^2/2-x^3/3-\ldots$ is appearing. Thus, the lowest order in the RPA correlation energy expansion  in $\chi(i\omega)V$ is given by the second-order,
\begin{align}\label{lp_2_freq}
    &E^{(2)}_c = -\frac{1}{4\pi}\int\limits_{0}^{\infty} d\omega \;\mathrm{Tr} 
    \{(\chi(i\omega)V)^2  \} 
    \\
&    =  -\frac{1}{4 }
    \sum_{\mu\mu'} \text{Tr}\{\chi_\mu V \chi_{\mu'} V\}\;\frac{1}{\pi}\int\limits_{0}^{\infty}  d\omega \;\phi_\omega(x_\mu)\,\phi_\omega(x_{\mu'}) 
    \,,\label{lp_2_freq_2}
\end{align}
which is precisely the direct 
second-order M\o ller-Plesset correlation energy (MP2)~\cite{haser1992laplace}. 
Inserting the Fourier transform~\eqref{susceptibility_cosine} of $\chi(i\omega)$ into Eq.~\eqref{lp_2_freq}, we obtain 
\begin{align}\label{lp_2}
    &E_c^{(2)} = -\frac{1}{2} \int\limits_0^{\infty} d\tau\;\mathrm{Tr}  \{(\hat{\chi}(i\tau) V)^2\}
    \\
    &=  -\frac{1}{4}
    \sum_{\mu\mu'} \text{Tr}\{\chi_\mu V \chi_{\mu'} V\}\;2\int\limits_{0}^{\infty} d\tau \;\hat{\phi}_\tau(x_\mu)\,\hat{\phi}_\tau(x_{\mu'})\label{lp_2_2}
    \,.
\end{align}
Eqs.~\eqref{lp_2_freq_2} and~\eqref{lp_2_2} evaluate to~\cite{kaltak2014low,haser1992laplace}
\begin{align}
\label{lp_2_analyt}
    E_c^{(2)} = -\frac{1}{4} \sum_{\mu\mu'}\,\text{Tr}\{\chi_\mu V \chi_{\mu'} V\}\,\frac{1}{x_\mu+x_{\mu'}}  \,.
\end{align}

\subsection{\label{sec:grids}Constructing minimax time and frequency grids}
Eqs.~\eqref{lp_2_freq}\,-\,\eqref{lp_2_analyt} are an ideal starting point to construct time and frequency grids.  
The frequency integral in  Eq.~\eqref{lp_2_freq_2} is discretized by a frequency grid
\begin{align}
\vv{\omega}=\{\omega_k\}_{k=1}^N
\end{align}
and  integration weights
\begin{align}
\vv{\gamma}=\{\gamma_k\}_{k=1}^N\,,
\end{align}
where $N$ is the number of integration points. 
Following Kaltak \textit{et al.}~\cite{kaltak2014cubic},
the grid generation is simplified by restricting Eqs.~\eqref{lp_2_freq_2} and \eqref{lp_2_analyt} to identical transition energies~$x_\mu\eqt x_{\mu'}$.
We require that the discretized frequency integral of Eq.~\eqref{lp_2_freq_2} is as close as possible to the exact result $1/(2x_\mu)$ from Eq.~\eqref{lp_2_analyt}, i.e.~the error function
\begin{equation}\label{f_err}
     \etafreq(x) =\frac{1}{2x}-\frac{1}{\pi}\sum_{k=1}^{N} \gamma_k\, \phi^2_{ \omega_k}(x)\,,   
\end{equation}
is minimized with respect to $\vv{\omega}$ and $\vv{\gamma}$.

Similarly, the time integral in the lower line of Eq.~\eqref{lp_2_2} 
is discretized by a time grid
\begin{align}
\vv{\tau}\eqt\{\tau_i\}_{i=1}^N
\end{align}
and  integration weights
\begin{align}
\vv{\sigma}\eqt\{\sigma_i\}_{i=1}^N\,.
\end{align}
The corresponding error function reads 
\begin{align}\label{t_err}
     \etatime(x) =\frac{1}{2x}-2\sum_{i=1}^{N} \sigma_i \hat{\phi}^2_{\tau_i}( x)\,.
 \end{align}
For both error functions, Eqs.~\eqref{f_err} and \eqref{t_err}, the transition energies are restricted to the interval
\begin{align}
x\in I^* &= \big[\emin, \emax\big]\, ,\\[0.5em]
\emin := \text{min}(\epsilon_{a}-\epsilon_{j})\; &, \hspace{1em}\emax := \text{max}(\epsilon_{a}-\epsilon_{j})\,,
\end{align}
where $a$, as before, refers to an unoccupied state, and $j$ to an occupied state.
The minimax grid parameters $\freqminimaxgrids$, $\timeminimaxgrids$ are defined as parameters that minimize the maximum norm of the error functions~$\etafreq(x)$ and~$\etatime(x)$, 
\begin{align}
 \freqminimaxgrids & := 
 \argminfreq\;\max_{x \in I^*}\;|\etafreq(x) |\,, \label{e13}
 \\[0.6em]
  \timeminimaxgrids & :=\argmintime\;\max_{x \in I^*}\;| \etatime(x)| \,.\label{e14}
\end{align}

It is convenient to consider minimax time and frequency grids  for the special interval~\cite{kaltak2014low,hackbusch2019computation} 
\begin{align}
x\in \tilde{I} := [1,R]\,,\hspace{2em} R := \frac{\emax}{\emin}\,.
\end{align}
The corresponding minimax grids $\freqminimaxgridstilde$ and $\timeminimaxgridstilde$ then only depend on $N$ and $R$,
\begin{align}
 \freqminimaxgridstilde & = 
 \argminfreq\;\maxxoneR|\etafreq(x) |\,, \label{e18}
 \\[0.6em]
  \timeminimaxgridstilde &  =\argmintime\;\maxxoneR| \etatime(x)| \,.\label{e19}
\end{align}
The minimax grids $\freqminimaxgrids$, $\timeminimaxgrids$ for a specific molecule or material with interval $x\intext I^* \eqt [\emin,\emax]$ easily follow by rescaling,~\cite{kaltak2014low,hackbusch2019computation} 
\begin{align}
\gks  &= \frac{\tgks}{\emin }\;,\hspace{3.1em} \wks = \frac{\twks}{\emin}\;,\label{e20}
\\[1em]
\sis &=  2\emin\tsis \;,\hspace{2em} \tis = 2\emin\ttis \,.\label{e21}
\end{align}
According to the alternation theorem of Chebychev, there exists a  reference set of $2N+1$ extrema points such that the maximum error of the quadrature is minimized. 
This property is used in the sloppy Remez algorithm~\cite{press2007numerical} to perform the minimizations in Eqs.~\eqref{e18} and ~\eqref{e19} in practice~\cite{kaltak2014low}.
\subsection{\label{sec:ct-st}Cosine and sine transformations}

To convert between imaginary time and frequency grids [Eqs.~\eqref{susceptibility_cosine}, \eqref{W_cosine}, \eqref{self-omega}], the functions $F(i\omega)$ and $\hat{F}(i\tau)$ are split into even and odd parts~\cite{liu2016cubic},
\begin{align}
 F(i\omega) &= \Fe(i\omega) + \Fo(i\omega)\,,\\[0.5em]  
 \hat{F}(i\tau) &= \Fhe(i\tau) + \Fho(i\tau)\,,  
 \end{align}
with $\Fe(x)\eqt \Fe(-x)$ and $\Fo(x)\eqt{-}\Fo(-x)$. 
The corresponding nonuniform discrete Fourier transforms turn into sine transforms and cosine transforms~\cite{liu2016cubic}, 
\begin{align}
    \label{ct_even}
    \Fe(i\omega_k) &= \sum_{j=1}^{N} \delta_{kj} \,\mathrm{cos}\,(\omega_k\tau_j)\Fhe(i\tau_j)\,,
    \\[0.5em]   
    \label{st_odd}
    \Fo(i\omega_k) &= i\sum_{j=1}^{N} \lambda_{kj} \,\mathrm{sin}\,(\omega_k\tau_j)\Fho(i\tau_j).
    \\[0.5em]       
    \label{ct_even_inv}
    \Fhe(i\tau_j) &= \sum_{k=1}^{N} \eta_{jk} \,\mathrm{cos}\,(\tau_j\omega_k)\Fe(i\omega_k).
     \\[0.5em]     
    \label{st_odd_inv}
    \Fho(i\tau_j) &= -i \sum_{k=1}^{N} \zeta_{jk} \,\mathrm{sin}\,(\tau_j\omega_k)\Fo(i\omega_k),
\end{align}
where the  time  points $\tau_j$ and  frequency points $\omega_k$ are precalculated from Eqs.~\eqref{e20} and~\eqref{e21}.
The Fourier weights $\delta_{kj}, \lambda_{kj}, \eta_{jk},\zeta_{jk}$ 
are computed by least-squares minimization in the following way, exemplarily shown for $\delta_{kj}$ from Eq.~\eqref{ct_even}: 
one inserts the auxiliary functions Eqs.~\eqref{axi_omega}, \eqref{axi_tau} for $\Fe(i\omega_k)$ and $\Fhe(i\tau_j)$ so that the error function of Eq.~(\ref{ct_even}) reads
\begin{eqnarray}\label{ct_min}
 \eta^c_{\overrightarrow{\delta\indk}}(x) = \frac{2x}{x^2+\omega_k^2}-\sum_{j=1}^N \delta_{kj}\,\mathrm{cos}(\omega_k \tau_j)\,e^{-x\tau_j}\,,
\end{eqnarray}
where we abbreviate $\overrightarrow{\delta\indk}\coloneqqt \{\delta_{kj}\}_{j=1}^N$.
 $\overrightarrow{\delta\indk}$ is then computed by linear-least-squares minimization of~$ \eta^c_{\overrightarrow{\delta\indk}}(x)$, 
\begin{align} 
\begin{split}
\overrightarrow{\delta\indk} &= ((\bX)^\text{T}\bX)^{-1}(\bX)^\text{T}\overrightarrow{y\indk}\,, 
 \\[0.5em]
  X_{\alpha j}\indk &= \mathrm{cos}(\omega_k \tau_j)e^{-x_\alpha\tau_j}\,,\hspace{1.2em}
 y_\alpha\indk = \frac{2x_\alpha}{x_\alpha^2+\omega_k^2} \,.
 \end{split}
 \label{e42}
\end{align}
The grid points~$\{x_\alpha\}$ are chosen in the interval $[1,R]$.

\subsection{\label{sec:eval} A priori assessment of the precision of minimax grids}

The sloppy Remez algorithm to generate minimax grids is numerically unstable and frequently finds only local minima and not the global optimization minimum.  In such cases, the minimax grid is not optimal for certain $N$ and certain $R$. However, not fully optimal minimax grids can be sufficiently accurate for  $GW$ or RPA.

In Sec.~\ref{sec:test}, we will describe detailed benchmarks for $GW$ and RPA calculations. However, before this, the quality of the grids
can already been assessed by testing exact properties of Fourier transforms. 
As a measure for the frequency integration error of the direct MP2 frequency integral in Eq.~\eqref{lp_2_freq} and the RPA frequency integral in Eq.~\eqref{eq:rpa}, we define the minimax error
\begin{align}
    \textfrak{d}(R,N) =\maxxoneR|\etafreqminimaxgrid(x) | \label{minimaxerror}\,.
\end{align}
$\etafreqminimaxgrid(x) $ is the error of the frequency integration and is defined in Eq.~\eqref{f_err}. $\freqminimaxgridstilde$ is the minimax grid as computed from Eq.~\eqref{e18}.

\begin{figure}[t]\centering   
\includegraphics[width=3in]{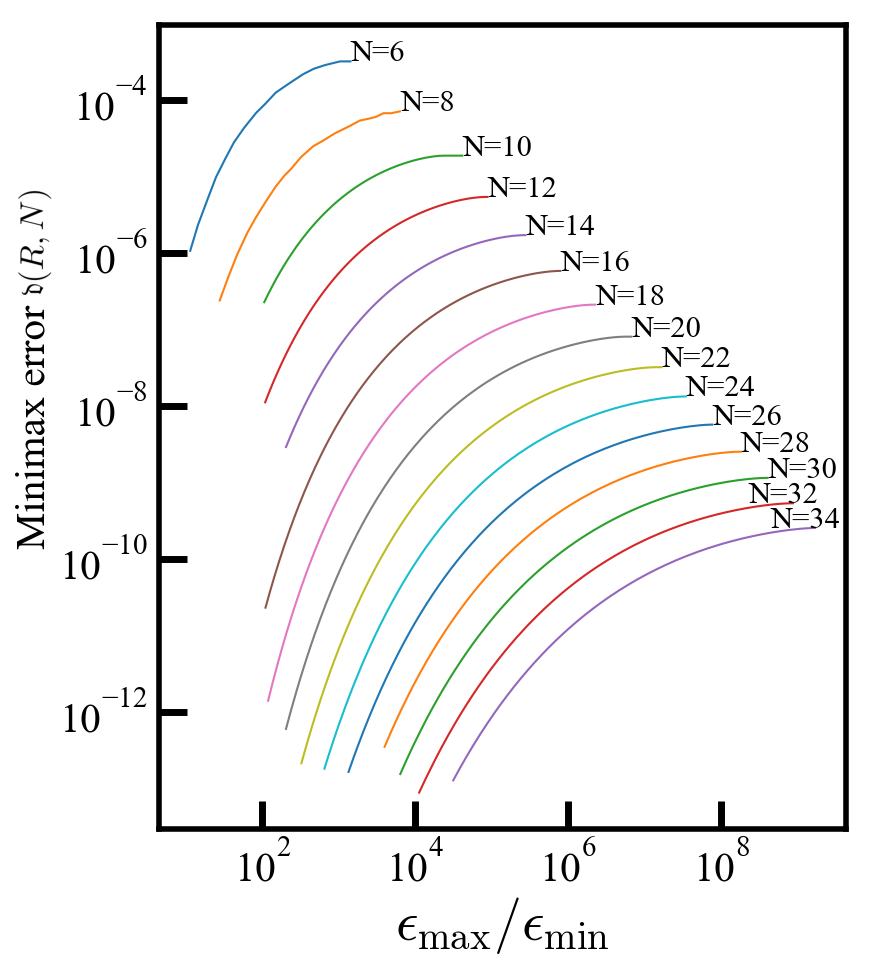} 
   \caption{Minimax error, Eq.~\eqref{minimaxerror}, as a function of the energy range $R\eqt\epsilon_\text{max}/\epsilon_\text{min}$ 
   for minimax grids with varying number of integrations points $N$.}
   \label{delta_N_Er}
\end{figure}

The minimax error~$\textfrak{d}(R,N)$ is plotted in Fig.~\ref{delta_N_Er} as a function of the number of grid points~$N$ and the energy range~$R$. We observe that for a given system (molecule or solid) with fixed range $R$, the error decreases exponentially with  increasing  $N$. Conversely, increasing $R$ by varying the system from small to large energy gaps leads to an increase in $\textfrak{d}$. An upper range~$R_c(N)$ exists, such that the minimax grid is identical for all $R \gt R_c(N)$ with saturating error,~$\textfrak{d}(R,N)\eqt \textfrak{d}(R_c(N),N)$ for $R \gt R_c(N)$~\cite{kaltak2014cubic,delben2015enabling}.
We thus only report minimax grids up to the upper  range~$R_c(N)$.

\section{Computational details}\label{computational_details}

\subsection{Test systems}\label{sec:testsystems}

For molecules, benchmarks were performed for the Thiel~\cite{Thielset} and GW100~\cite{van2015gw} sets. 
The Thiel set consists of 28 small organic molecules composed of C, N, O and H. The  geometries were taken from Ref.~\citenum{Thielset}. 
The GW100 set contains small molecules with covalent and ionic bonds covering a wide range of the periodic table. The geometries were taken from Ref.~\citenum{van2015gw}. 
For 2D semiconductors, four monolayer materials MoS$_2$, MoSe$_2$, WS$_2$ and WSe$_2$
were considered. The geometries were taken from the C2DB database~\cite{Gjerding2021recent}.
For three-dimensional (3D) bulk crystals, the diamond structure of Si and C, zinc-blende structure of BN, GaAs and SiC, and the rocksalt structure of MgO and LiF were selected. The lattice parameters were taken from Ref.~\citenum{liu2016cubic}. The \textit{k}-mesh settings given in Table~\ref{abinit_benchmark_test} were used for all calculations. In the interest of open materials science data \cite{Himanen/Geurts/Foster/Rinke:2019,Ghiringhelli2023}, we made the input and output files of all calculations available on the NOMAD repository~\cite{nomad_repoGW100_lowscaling}.

\subsection{FHI-aims}

FHI-aims is an all-electron electronic structure code based on numerical atom-centered orbitals (NAOs) \cite{Blum2009}. 
All canonical RPA calculations were performed with FHI-aims. The RPA implementations in FHI-aims scale $\mathcal{O}(N_\text{at}^4)$ with respect to system size $N_\text{at}$ and are based on the resolution-of-the-identity (RI) approach,~\cite{RI_Whitten73,RI_Dunlap79,RI_Mintmire82,RI_Vahtras93} refactoring the 4-center electron repulsion integrals in a product of three- and two-center integrals. For finite systems, the RPA implementation~\cite{Ren2012} relies on a global RI scheme with a Coulomb metric (RI-V), while for extended systems a localized variant (RI-LVL) is employed~\cite{Ihrig_2015,Ren/etal:2021}.

Molecular RPA calculations were performed for the Thiel benchmark set using the cc-pVTZ Dunning basis set~\cite{ccpvtzbasis}, which is an all-electron basis set of contracted spherical Gaussian orbitals. In FHI-aims, Gaussian basis sets are presented numerically to adhere to the NAO scheme. For the crystalline systems, we employed the hierarchical \textit{tier 2}  NAO basis functions~\cite{Blum2009}.
For both, molecules and crystal, the RI auxiliary basis functions (ABFs) were auto-generated during run-time following the approach described in Ref.~\citenum{Ren2012}. For the crystalline systems, we have added additional \textit{4f} and \textit{5g} functions to the auto-generated ABFs to increase the accuracy of the RI-LVL approach implemented for solids ~\cite{Ren/etal:2021}. 

The RPA correlation energies were numerically computed on imaginary frequency grids~\cite{Ren2012}. We benchmarked three different grid types: the Gauss-Legendre (GL), modified Gauss-Legendre (mod-GL) and minimax quadrature grids. The modified Gauss-Legendre quadrature is described in more detail in Ref.~\citenum{Ren2012}. 

\subsection{CP2K}

CP2K is an electronic-structure code that employs Gaussian basis sets for expanding the KS orbitals~\cite{Kuehne2020}.  CP2K can be used with Goedecker-Teter-Hutter  pseudopotentials~\cite{Goedecker1996} (GTH) or as an all-electron code using the Gaussian and augmented plane-waves  scheme (GAPW)~\cite{Lippert1999}. CP2K features a low-scaling $G_0W_0$ implementation based on a local-orbital-basis adaptation of the space-time method~\cite{rojas1995space}, where sparsity is introduced by combining a global RI scheme with a truncated Coulomb metric~\cite{wilhelm2021low,graml2023lowscaling}. Low-scaling $GW$ implementations for finite systems~\cite{wilhelm2021low} and  recently also for extended systems~\cite{graml2023lowscaling} are available.

We used the low-scaling $G_0W_0$ implementations in CP2K for performing the all-electron GW100 benchmark~\cite{van2015gw} and for computing the $G_0W_0$ bandgap of 2D materials with pseudopotentials. For the GW100 benchmark, we expanded molecular orbitals in the all-electron def2-QZVP~\cite{Weigend2003} basis set and  used def2-TZVPPD-RIFIT~\cite{Haettig2005} as ABFs.
For $G_0W_0$ benchmark calculations on 2D materials, we used 10\,$\times$\,10 supercells, GTH pseudopotentials~\cite{Goedecker1996} with a TZV2P-MOLOPT basis set~\cite{VandeVondele2007} and corresponding ABFs~\cite{graml2023lowscaling}.
For molecules and 2D materials, we set a truncation radius  of 3\,\AA~\cite{Vahtras1993,Jung2005} for the truncated Coulomb metric.
Two- and three-center Coulomb integrals over Gaussian basis functions were computed with analytical schemes~\cite{Golze2017,libint}.  
The self-energy was analytically continued from the imaginary to the real-frequency domain using a Pad\'{e} model~\cite{Vidberg1977,van2015gw} with 16 parameters.

\subsection{$\mathrm{ABINIT}$}\label{abinit_comp_details}
ABINIT is an electronic-structure code that relies on plane-waves for the representation of wavefunctions, density, and other space-dependent quantities, with pseudopotentials or projector-augmented waves~\cite{Gonze2020a, Gonze2020b}. 
We used ABINIT for the calculations of $G_0W_0$ band-gaps of the crystalline systems with the specifications given in Table~\ref{abinit_benchmark_test} and norm-conserving pseudopotentials~\cite{Hamann2013} from the PseudoDojo project~\cite{Setten2018} (standard accuracy table and the recommended cutoff energy for the plane-wave expansion of the KS states). 
\begin{table}[h]
\centering
      \caption{Parameters for periodic conventional $GW$ and low-scaling  $GW$ calculations when using $\mathrm{ABINIT}$. $E_c$ denotes the energy cutoff for expanding Bloch wavefunctions, $E_c^\epsilon$ the energy cutoff for the dielectric matrix and $N_\text{bands}$ the number of bands for computing the Green's functions.
      All the $k$-meshes are $\Gamma$-centered.}
        \centering
         \begin{ruledtabular}
        \begin{tabular}{l c c c c c c c c}
          & & $k$-mesh  & &  $E_c$  (Ha)& & $E_c^\epsilon$ (Ha)& &$N_\text{bands}$ \\
         \hline
          Si  & & 4 $\times$ 4 $\times$ 4 & & 24 & &10 & & 500\\
          LiF  & & 8 $\times$ 8 $\times$ 8 & & 48 & &12 & & 1000\\
          SiC  & & 8 $\times$ 8 $\times$ 8 & & 48 & &12 & & 1000\\
          C  & & 6 $\times$ 6 $\times$ 6 & & 45 & &8 & & 1000\\
          BN  & & 4 $\times$ 4 $\times$ 4 & & 48 & &12 & & 1000\\
          MgO  & & 4 $\times$ 4 $\times$ 4 & & 50 & &12 & & 1000\\
          GaAs  & & 8 $\times$ 8 $\times$ 8 & & 48 & &12 & & 1000\\
        \end{tabular}
         \end{ruledtabular}
        \label{abinit_benchmark_test}
\end{table}
The conventional $GW$ implementation in ABINIT scales with $O(N_{\text{at}}^4)$ and computes the RPA susceptibility along the imaginary frequency axis using the exact Adler-Wiser expression. For the conventional $GW$ calculations, the self-energy was evaluated directly along the imaginary frequency axis using a linear mesh extending up to 50 eV and 60 points. 
The new cubic-scaling \textit{GW} implementation is based on the space-time approach~\cite{rojas1995space} and will be made available in ABINIT version 10.
In the low-scaling $GW$ calculations, we use minimax grids with 10, 20 and 30 points provided by the GreenX library~\cite{GreenX}. 
The non-interacting susceptibility and the correlation part of the self-energy were computed using Fast Fourier Transforms in a supercell followed by a Pad\'e-based analytic continuation of the matrix elements of the self-energy as discussed in Ref.~\citenum{liu2016cubic}. In all conventional and low-scaling $GW$ calculations, the number of imaginary frequencies used for the Pad\'e was set to be equal to the number of frequency points used to sample the self-energy.
 
The integrable Coulomb singularity 
is treated by means of the auxiliary function
proposed in Ref.~\citenum{Carrier2007} to accelerate the convergence with respect to Brillouin zone sampling.
The calculation of the head and wings of the polarizability takes into account the non-local part of the KS Hamiltonian that arises from the pseudopotentials into account~\cite{Baroni1986}.

\section{\label{sec:test}\textit{GW} and RPA benchmark calculations with minimax grids from GreenX}

\subsection{\label{subsec:fhi_aims} Conventional  RPA for molecules and solids}

\begin{figure}[t]
    \centering
    \includegraphics[width=0.45\textwidth]{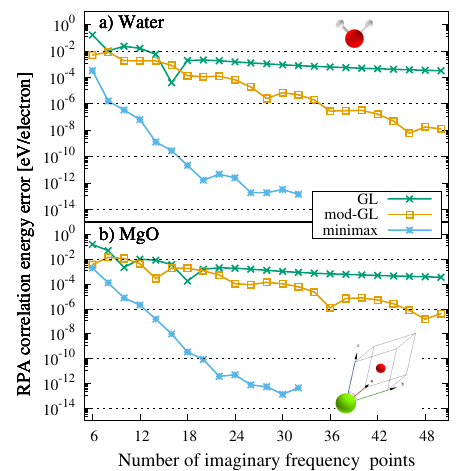}
    \caption{Errors of the RPA correlation energy for the water molecule (top) and the MgO bulk (bottom) using the GL, mod-GL and minimax imaginary frequency grids. The RPA error reported on the vertical axis employs as reference point the RPA energy calculated with the minimax grid containing 34 points. The error is given in eV/electron, dividing the absolute RPA correlation energy by the number of electrons per molecule or by the number of electrons in the unit cell.}
    \label{rpa_ener}
\end{figure}

\begin{figure}[t]
    \centering
    \includegraphics[width=0.45\textwidth]{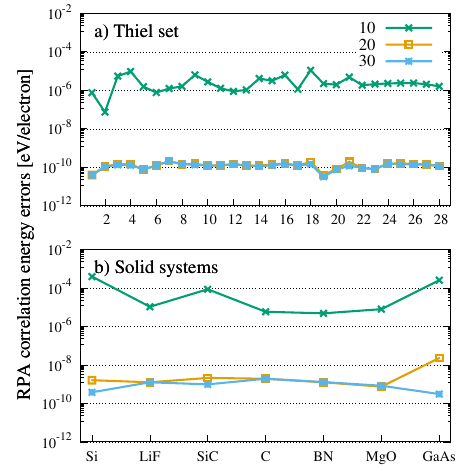}
    \caption{Errors of the RPA correlation energy of a) the Thiel set and b) solids when using  minimax frequency grids with 10, 20 and 30 grid points.
    The reference for computing the RPA error is the mod-GL grid with 200 frequency grid points. The error is given in eV/electron, dividing the absolute RPA correlation energy by the number of electrons per molecule or by the number of electrons in the unit cell.}
    \label{rpa_energies_set}
\end{figure}
While the primary purpose of the minimax grids is to facilitate  low-scaling RPA and $GW$ calculations, they can be also utilized to reduce the computational prefactor in conventional RPA implementations with $O(N_{\text{at}}^4)$ complexity. We demonstrate this in Fig.~\ref{rpa_ener} for a single water molecule  (top) and bulk MgO (bottom). Figure~\ref{rpa_ener} reports the  error of the RPA correlation energies calculated with GL and mod-GL grids with 6–50 points and minimax grids with 6–32 points. The RPA energy obtained with the 34-point minimax grid is used as reference. 

Figure~\ref{rpa_ener} shows that the GL quadrature converges slowly, while the mod-GL grids exhibit a faster convergence as already discussed in Ref.~\citenum{Ren2012}. The minimax grids display the fastest convergence. We set a target accuracy of $10^{-6}$~eV/electron, which is motivated by the size of the error introduced by the RI-V approach. RI-V is implemented in FHI-aims for finite RPA calculations and considered the most accurate RI scheme~\cite{Vahtras1993}. Recently, we showed for MP2 correlation energies that RI-V introduces an average error of $5\times 10^{-6}$~eV/electron compared to RI-free results~\cite{Delesma2024}. Similar RI errors are expected for RPA. The GL grids fail to achieve the desired accuracy for the grid sizes presented in Fig.~\ref{rpa_ener}. In contrast, the mod-GL grids surpass the accuracy threshold with over 36 and 46 frequency points for the water molecule and bulk MgO, respectively. With minimax grids, errors below $10^{-6}$ eV/electron are already achieved with 10 frequency points for water and 14 for MgO. Minimax frequency grids are therefore approximately 3.5-fold more efficient than mod-GL grids.

For even larger minimax grids of 20 points, the RPA integration errors drop below $10^{-10}$\,eV/electron in both cases, as is evident in Fig.~\ref{rpa_ener}. The minimax error further decreases by two orders of magnitude to $10^{-12}$~eV/electron before leveling out at around 32 points. We found that it is not possible to converge the RPA correlation energies as tightly with the mod-GL grids, which eventually yield fluctuating errors around $10^{-9}$ - $10^{-10}$~eV/electron. This is also the reason why we use the 34-point minimax grid as reference in Fig.~\ref{rpa_ener} instead of a mod-GL grid with hundreds of points. The convergence behavior of the minimax grids for more than 20 grid points would not be adequately displayed.

Further benchmarks were performed with 10, 20 and 30 minimax grid points for molecules and crystalline systems as shown in Figs.~\ref{rpa_energies_set}\,(a) and (b). Taking the mod-GL result with 200 frequency points as our reference, we find that minimax grids with 10 points fail to converge all molecular and extended systems below the threshold of the RI-V accuracy. However, we observe an excellent convergence for larger grids. All molecules from the Thiel set are converged within $10^{-9}$~eV/electron for minimax grids with 20 or more frequency points; see Fig.~\ref{rpa_energies_set}\,(a).

For solids shown in Fig.~\ref{rpa_energies_set}\,(b), we observe that the RPA correlation energy is converged to better than $10^{-8}$\,eV/electron for 30 minimax frequency points. For 20 minimax points, the RPA correlation energy of GaAs still deviates by $2.2\cdot10^{-8}$\,eV/electron from the reference. This error is reduced by two orders of magnitudes for 30 minimax points instead. We note that even an error of $2.2\cdot10^{-8}$~eV/electron is excellent when taking the RI-V error as target accuracy. Minimax grids with 20 grid points can therefore be considered a reliable choice for conventional RPA calculations.
For completeness, the RPA data for the four systems (or set of systems), for the GL, modified GL and minimax cases, are gathered in the Supplemental Material \cite{supp} (Tables S1 to S4).

\subsection{\label{subsec:GW100}Low-scaling \textit{GW} for molecules}

\begin{figure}[h!]
    \centering
    \includegraphics[width=0.45\textwidth]{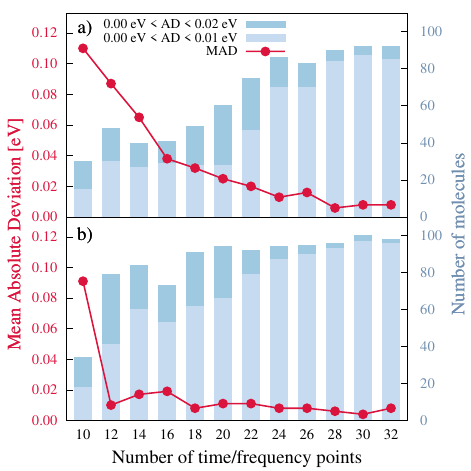}
    \caption{$GW100$ benchmark of the $G_0W_0$@PBE energies computed with the low-scaling algorithm for a) HOMOs and b) LUMOs. The bars represent the number of molecules with a given absolute deviation from the FHI-aims values reported in  Ref.~\citenum{van2015gw}. The mean absolute deviation (MAD) as a function of number of time/frequency points is given by the red dots. Table~S5 in the Supplemental Material \cite{supp} reports the raw data.}
    \label{GW100_AD_MAD}
\end{figure}

In this section, we present GW100 benchmark results for our minimax time- and frequency grids employing the low-scaling $G_0W_0$ implementation in CP2K~\cite{wilhelm2021low}. We computed quasiparticle energies of the highest occupied molecular orbitals (HOMOs) and lowest unoccupied molecular orbitals (LUMOs), for minimax grids with 10-32 points. We selected as our reference the $G_0W_0\mathrm{@PBE}$ results reported in Ref.~\citenum{van2015gw}, more precisely, the analytic continuation results from FHI-aims based on the Pad\'e model and the def2-QZVP basis set. For the HOMOs, the five multi-solution cases were excluded, namely, BN, BeO, MgO, O$_3$ and CuCN. 

Figure~\ref{GW100_AD_MAD} summarizes the convergence of the $G_0W_0\mathrm{@PBE}$ HOMO and LUMO energies with respect to the number of minimax time and frequency points. 
We find that the mean absolute deviation (MAD) steadily decreases with the number of minimax points. 
Already for 24 minimax points, we observe an MAD smaller than 20\,meV for both HOMOs and LUMOs. For grids with 28 points, the MAD drops below 10\,meV. Figure~\ref{GW100_AD_MAD} additionally reports the number of molecules with an absolute deviation (AD) below 10~meV and 20~meV. For 32 minimax points, 85 (out of 95) HOMO energies and 97 (out of 100) LUMO energies deviate by less than 10~meV from the reference value. Moreover, 92 HOMO and 100 LUMO energies are within the threshold of 20~meV.

For molecular benchmarks, our GW100 results are well within the target accuracy of a few meV we aim for when comparing different $GW$ implementations using the same basis set. For example, the original GW100 benchmark study~\cite{van2015gw} reported a MAD of 3~meV and 6~meV for the HOMO and LUMO, respectively, comparing FHI-aims and RI-free Turbomole results at the def2-QZVP level.
Furthermore, we conducted a similar convergence study as shown in Fig.~\ref{GW100_AD_MAD} in our previous work,~\cite{wilhelm2021low} where we also reported MADs below 10~meV for minimax grid sizes $\ge28$ points.

\begin{figure}[t]
    \centering
    \includegraphics[scale=1]{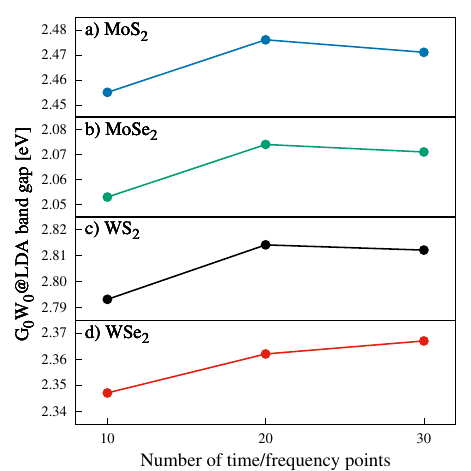}
    \caption{$G_0W_0$@LDA bandgap of monolayer semiconductors computed with  CP2K~\cite{graml2023lowscaling,Kuehne2020} as function of the number of   points of the minimax time-frequency grid. Table~S6 in the Supplemental Material \cite{supp} reports the raw data.
    }
    \label{GW_2D_mater}
\end{figure}

\subsection{Low-scaling \textit{GW} for 2D materials}

The unit cells of 2D materials can become large quickly, possibly requiring a hundred to thousands of atoms. An example are  Moir\'e structures formed from twisted transition metal dichalcogenide (TMDC) bilayers~\cite{Shabani2021,Mak2022,Huang2022}. The development of low-scaling algorithms is a promising strategy to make $GW$ calculations for these systems computationally feasible~\cite{graml2023lowscaling}. Localized-basis-set frameworks are here particularly advantageous due to the large vacuum regions which need to be added below and above the 2D slabs~\cite{qiu2016screening}. Such vacuum regions drastically increase the computational cost for plane-wave implementations, but not for localized-basis-functions schemes.

In this section, we present $G_0W_0$@LDA bandgap calculations for monolayer MoS$_2$, MoSe$_2$, WS$_2$, and WSe$_2$, which have previously been used as building blocks for TMDC-based Moir\'e structures~\cite{Mak2022,Huang2022}. Figure~\ref{GW_2D_mater} reports the band gaps computed with 10, 20 and 30 time and frequency points using the recently developed periodic, low-scaling $GW$ implementation in CP2K~\cite{graml2023lowscaling}. We find that the $G_0W_0$@LDA band gap changes on average by 20\,meV and when increasing the minimax mesh size from 10 to 20 points and by 4\,meV from 20 to 30 points. These findings indicate that a grid size of 20 is a reliable choice for the 2D case.

We report absolute gaps in Fig.~\ref{GW_2D_mater} since reliable reference data for 2D materials are generally difficult to find~\cite{golze2019gw}. Qiu \textit{et al.}~\cite{qiu2016screening} showed that a very fine $k$-grid sampling is required for TMDC monolayers ($24\times24\times1$ grids). Equivalently, our recent work~\cite{graml2023lowscaling} demonstrated a slow convergence with the supercell size, necessitating supercell sizes of at least $10\times10\times1$. Due to these challenges, a benchmark set for highly accurate $GW$ band structures of 2D monolayers has not been established yet to the best of our knowledge. 
In previous work~\cite{graml2023lowscaling}, well-converged $G_0W_0$@LDA gaps were collected from several codes~\cite{Rasmussen2021,Camarasa2023,graml2023lowscaling} for the MoS$_2$, MoSe$_2$, WS$_2$, and WSe$_2$ monolayers, observing an average deviation of 55~meV between the codes. The accuracy of a minimax time-frequency grid with 10 points is already better than the difference between different $GW$ codes. 

\subsection{\label{subsec:abinit_results}Low-scaling \textit{GW} for solids}
 
As a preliminary accuracy check of the minimax grids, we have compared selected matrix elements of the susceptibility in Fourier space computed along the imaginary axis for conventional and low-scaling $GW$ implementations (see Fig.~S1 in the Supplemental Material \cite{supp}). The agreement between the two implementations is excellent. Small differences only arise for  Fourier components of the imaginary part (see middle-right panel of Fig.~S1 in the Supplemental Material \cite{supp}) but this is essentially numerical noise that does not affect our final results. 

\begin{figure}[h!]
  \centering
    \includegraphics[scale=1.0]{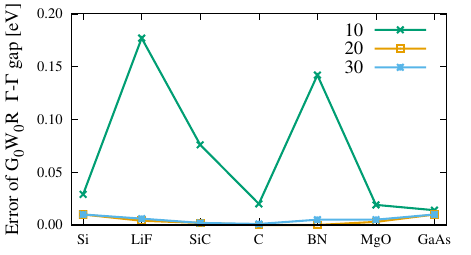} 
  \caption{The low-scaling $GW$ error with respect to the conventional $GW$ using ABINIT (See Table~S7 in the Supplemental Material \cite{supp} for the raw data).}
  \label{gwr_error_abinit}
\end{figure}

We then computed  the direct band gaps of the seven solids introduced in Section~\ref{sec:testsystems} with the low-scaling $GW$ implementation in ABINIT and then compared with the analogous results obtained with the conventional quartic-scaling $GW$ implementation. The results obtained with the settings given in Table~\ref{abinit_benchmark_test} are summarized in Fig.~\ref{gwr_error_abinit}. As becomes  clear from Fig.~\ref{gwr_error_abinit}, the band gaps of the low-scaling implementation approach those of the conventional one with increasing number of frequency points. For 20 minimax points the difference is already below 10~meV. Our results are in line with a previous comparison for these materials. Liu \textit{et al.}~\cite{liu2016cubic} also found a difference of 10~meV between their low-scaling and the conventional quartic-scaling $GW$ implementation~\cite{Shishkin2006} in the VASP code, using 20 minimax grid points in their space-time-based low-scaling implementation.

\section{\label{sec:sum}Conclusion}

The time-frequency component of the GreenX library~\cite{Azizi2023}, which provides minimax time-frequency grids along with corresponding quadrature weights to compute correlation energies and quasiparticle energies and weights for Fourier transforms, has recently been released. In the present work, we linked the GreenX library to three codes (FHI-aims, CP2K and ABINIT), and performed conventional RPA and low-scaling $GW$ calculations for a variety of systems. Our test systems include the molecules of the Thiel and GW100 benchmark sets, four 2D-dimensional semiconductors (TMDC monolayer materials) and seven 3D bulk crystals. We found that the conventional RPA calculations are well converged within $<10^{-7}$~eV/electron for minimax grids with 20 points, reducing the computational prefactor by a factor of $\approx 3$ compared to calculations with conventional modified Gauss-Legendre grids. The RPA integration errors decrease to $<10^{-8}$~eV/electron for even larger grids with 30 points. Low-scaling $GW$ quasiparticle energies are converged to $\le10$~meV for all systems using only 30 minimax points. In most cases this excellent accuracy is already reached with 20 minimax grid points. 

Our findings show that the time-frequency component of GreenX provides a reliable foundation for the development of low-scaling RPA and $GW$ algorithms based on the space-time method. It also demonstrates its suitability to improve the computational efficiency of conventional RPA calculations.

\section{\label{sec:ack} ACKNOWLEDGMENTS}
M. A. acknowledges helpful discussions with Bogdan Guster. 
This work has been supported by the European Union’s Horizon 2020 research and innovation program under the grant agreement No 951786 (NOMAD CoE) and the Horizon Europe MSCA Doctoral Network Grant No. 101073486 (EUSpecLab). 
J.~W. and D.~G. acknowledge funding by the Deutsche Forschungsgemeinschaft (DFG, German Research Foundation) via the Emmy Noether Programme (Project No. 503985532 and 453275048, respectively).
The authors gratefully acknowledge the computing time provided to them on the high performance computer Noctua 2 at the NHR Center PC2. These are funded by the Federal Ministry of Education and Research and the state governments participating on the basis of the resolutions of the GWK for the national high-performance computing at universities (www.nhr-verein.de/unsere-partner).
The authors wish to acknowledge CSC – IT Center for Science, Finland, for computational resources.
The computational resources provided by  Aalto Science-IT are also acknowledged.
Computational resources
have been provided by the supercomputing facilities
of the Université catholique de Louvain (CISM/UCL),
and the Consortium des Equipements de Calcul Intensif en
Fédération Wallonie Bruxelles (CECI) funded by the
FRS-FNRS under Grant No. 2.5020.11. 

\medskip
\bibliography{refs.bib} 

\end{document}